\documentstyle[12pt]{article}
\newcommand{\sect}[1]{\setcounter{equation}{0}\section{#1}\indent}
\renewcommand{\theequation}{\thesection.\arabic{equation}}
\renewcommand{\thefootnote}{\fnsymbol{footnote}}
\newcommand{\EQ}{\begin{equation}}
\newcommand{\EN}{\end{equation}}
\newcommand{\bea}{\begin{eqnarray}}
\newcommand{\ena}{\end{eqnarray}}
\newcommand{\vs}[1]{\vspace{#1 mm}}

\newcommand{\e}{\epsilon}

\newcommand{\pa}{\partial}

\newcommand{\uda}{\nearrow \kern-1em \searrow}

\newcommand{\la}{\lambda}
\newcommand{\La}{\Lambda}
\newcommand{\dis}{\displaystyle}
\newcommand{\ti}{\tilde}

\makeatletter
\def\eqnarray{%
 \stepcounter{equation}%
 \let\@currentlabel=\theequation
 \global\@eqnswtrue
 \global\@eqcnt\z@
 \tabskip\@centering
 \let\\=\@eqncr
 $$\halign to \displaywidth\bgroup\@eqnsel\hskip\@centering
 $\displaystyle\tabskip\z@{##}$&\global\@eqcnt\@ne
 \hfil$\displaystyle{{}##{}}$\hfil
 &\global\@eqcnt\tw@$\displaystyle\tabskip\z@{##}$\hfil
 \tabskip\@centering&\llap{##}\tabskip\z@\cr}
\makeatother

\begin{document}

\begin{titlepage}
\setcounter{page}{0}
\begin{flushright}
EPHOU 96-008\\
December 1996\\
\end{flushright}

\vs{6}
\begin{center}
{\Large On Explicit Evaluations Around 
the Conformal Point in \\$N=2$ Supersymmetric Yang-Mills Theories
}

\vs{6}
{\large
Takahiro
Masuda
\footnote{e-mail address: masuda@phys.hokudai.ac.jp}
\\ and \\
Hisao Suzuki\footnote{e-mail address: hsuzuki@phys.hokudai.ac.jp}}\\
\vs{6}
{\em Department of Physics, \\
Hokkaido
University \\  Sapporo, Hokkaido 060 Japan} \\
\end{center}
\vs{6}

\centerline{{\bf{Abstract}}}
We show how to give the expression for periods, 
Higgs field and its dual of $N=2$ supersymmetric Yang-Mills theory 
 around the conformal point. This is 
achieved by evaluating the integral 
representation in the weak coupling region, and by using 
analytic continuation to 
the conformal point. 
The explicit representation is shown for 
the $SU(2)$ theory with matter fields and 
also for pure $SU(N)$ and pure $SO(2N)$ theory around the conformal point 
where the relation to the beta function of the theory is clarified. 
 We also discuss a relation between the fixed points 
in the $SU(2)$ theories with matter fields and 
the Landau-Ginzburg point of 2-D $N=2$ SCFT. 
\end{titlepage}
\newpage

\renewcommand{\thefootnote}{\arabic{footnote}}
\setcounter{footnote}{0}

\sect{Introduction} 

Recent years many progress about four dimentional 
$N=2$ supersymmetric Yang-Mills theories have been made.
 Seiberg and Witten \cite{SW} solved the low energy effective theory 
in the $SU(2)$ theory without matter fields exactly,  
 based on the duality and holomorphy by introducing the 
elliptic curve. Following this work various generalizations 
introducing the matter fields or gauge group being higher than $SU(2)$  
have been investigated by 
many people \cite{KLTY,APS,HO,DS,Hanany,AS,MW,CDF}. 
The solution of these theories
 extesively studied in the weak coupling region by various method, 
such as solving the Picard-Fuchs 
equation \cite{KLT,IY}, purtubative treatments to obtain the 
prepotential \cite{DKP,Ohta}.
The direct tests have been made by comparison with the 
instanton method in the case of $SU(2)$ theory with matter fields
\cite{IS1,HS}, and even for the higher rank gauge groups \cite{IS2}, 
 which provide the consistent results. 

Apart from the analysis in the weak coupling region, the power of the 
exact results should be used in the analysis in the strong 
coupling region, where one finds truely non-perturbative results. 
 Among the analysis of the strong coupling region, one of the 
striking fact is the existence of the conformal  
points \cite{AD,APSW,EHIY} where the prepotentials have no dependence 
of the dynamical mass scale. These theories are classified by 
 scaling behaviors around the conformal point \cite{EHIY}. 
 It seems interesting to investigate the theories around 
this points by deriving the explicit form of fields, 
which seems to provide us of a more concrete 
behaviour of the critical theories. 

In previous works \cite{MS1,MS2} 
we have evaluated the integral representation about the period, 
Higgs field and its dual on such situations. 
The results for  $SU(2)$ Yang-Mills theory 
with matter fields \cite{MS1} can  be 
analytically continuated  around the conformal point 
when the bare masses take the critical values. 
In this article, we generalize this approach 
to investigate the expression around the conformal point.  
We treat moduli parameters and bare masses 
as deviations from the conformal point.  
After evaluating the integral representation explicitely 
in the region where only one parameter is very large but 
other parameters are near the confomal point, we perform 
the analytic continuation of one large parameter to be near 
the conformal point. By use of the analyic continuation we can get the 
expression around the conformal point. 
 Usually, analytic continuation to the region where 
the logarithmic singularity exits must be treated with care because 
the result may depend on the choice of variables. 
In other words, some choice 
of variable are valid only within some branch. However, when 
we consider the analytic continuation to the 
critical point where there is no such singularity, the confirmation of
 such analytic continuation turns out to be
 easy, as we will see in each cases. 
As the matter of 
fact, this 
approach can be considered as 
 a generalization of the one given to obtain the periods 
for Calabi-Yau systems \cite{BCOFHJQ}. Of course, there are a variety of 
the class of conformal point \cite{EHIY} so that  we cannot 
exhaust all known cases. In this paper, 
we will deal with $SU(2)$ Yang-Mills theory 
with massive hypermultiplets, and $SU(N)$ and also 
 $SO(2N)$ Yang-Mills theory without matters. 
 We also provide expressions of Higgs field and its dual around 
the conformal point for the 
pure $SU(N)$ and also $SO(2N)$ Yang-Mills theories. 

This article is organized as follows. 
In section 2, we will obtain the expression of the fields around 
conformal points in $SU(2)$ theory 
 with matter fields case, 
and verify that the result recovers our previous 
result which was obtained by transformations 
of the hypergeometric functions\cite{MS1} 
when the deviation of mass parameters from the 
conformal point are set to be zero. We will also discuss the 
relation to 2-D $N=2$ SCFT through the simple correspodence to 
the deformation of curve on $W\bf{CP}^2$. This relation has been  
pointed out recently in the different context in the ref. \cite{LMW}.  
In section 3, 
we will derive the form of Higgs field and its dual in 
 the pure $SU(N)$ theory around 
the conformal point and clarify the relation to the beta function 
of theory. In section 4, we will study the pure $SO(2N)$ theory 
around the conformal 
point and discuss the validity of the expression. 
The last section will be devoted to some discussion.

\sect{$SU(2)$ Yang-Mills theories with matter fields}

In this section we treat the $SU(2)$ theory with $N_f$ matter fields ($N_f\le 
3$), to verify that our approach recover the previous results which 
was obtained by transformations of the hypergeometric functions\cite{MS1}. 

First of all we consider $N_f=1$ case, whose
 curve of $N_f=1$ theory is given by 
\bea
y^2&=&x^2(x-u)+{1\over 4}m\La^3 x-{\La^6\over 64}.
\ena
In order to calculate the periods:  
\bea
{\pa a\over \pa u}=\oint_{\alpha}{dx\over y},\ \ \ 
{\pa a_D\over \pa u}=\oint_{\beta}{dx\over y},
\ena
around the conformal point $u={3\over 4}\La^2,\ 
m={3\over 4}\La$ where 
the curve becomes degenerate as $y^2=(x-{\La^2\over 4})^3$,
we introduce the deviations from this conformal 
point as $\ti u=u-{3\over 4}\La^2,\ \ti m=
m-{3\over 4}\La$. 
The strategy is that after calculating the period in the weak coupling region
$\ti u\sim \infty$, we analytically continuate around 
the conformal point $\ti u\sim 0$. It should be noted that 
similar consideration has been used to evaluate periods for 
Calabi-Yau manifolds \cite{BCOFHJQ}. 
Rescaling the variable as $x={1\over 4}\La^2 z$ the curve becomes 
\bea
y^2=({\La^2\over 4})^3(z-1)^3-({\La^2\over 4})^2\ti u z^2-({\La^2\over 4})^2
\La mz,
\ena
and 
thus we consider $\ti u\over \La^2$ and 
$\ti m\over \La$ as perturbations from the conformal point. 
Expanding the period with respect to $1/\ti u$ 
we have the expression for the period as follows:
\bea
\oint{dx\over y}=
\ti u^{-1/2}\int_{-i\infty}^{i\infty}{ds\over 2\pi i}
 \sum_{m=0}^{\infty}& &{\Gamma({1\over 2}+s)
\Gamma(s+1)\Gamma(-s)\over \Gamma({1\over 2})\Gamma(s-m+1)m!
}\nonumber \\
& &\times \oint dz\, (1-z)^{3(s-m)}z^{-2s+m-1}\left(
{\La^2 \over 4\ti u}\right)^s\left({\ti m\over \La}\right)^m, 
\ena
where we have introduced Barnes-type integral representation. 
 From this expression we can 
find that ${\pa a\over \pa u}$ is 
obtained by picking up poles $z=0$ along $\alpha$ cycle 
in the weak coupling region. 
In this way, 
we find ${\pa a \over \pa u}$ in the weak coupling region is of the form
\bea
{\pa a\over \pa u}&=&{\sqrt 3\over \ti u^{1/2}2\pi}
\sum_{n,m=0}^{\infty}{\Gamma(n+m+{1\over 2})\Gamma(n+{1\over 3})
\Gamma(n+{2\over 3})\over \Gamma(n+{m\over 2}+{1\over 2})
\Gamma(n+{m\over 2}+1)n!m!}
\left(-{27\La^2\over 16\ti u}\right)^n
\left({\La\ti m\over 8 \ti u}\right)^m\nonumber \\
&=&{\sqrt 3 \over \ti u^{1/2}2\pi}\sum_{m=0}^
{\infty}{\Gamma(m+{1\over 2})\Gamma({1\over 3})\Gamma({2\over 3})\over 
\Gamma({m+1\over 2})\Gamma({m\over 2}+1)m!}
\left({\La\ti m\over 8 \ti u}\right)^m\\
& &\hspace{4cm}\times \, _3F_2\left({1\over 3},
{2\over 3},{1\over 2}+m;{m+1\over 2}
,{m\over 2}+1;-{27\La^2\over 16\ti u}\right),\nonumber
\ena
where $_3F_2$ is 
 the generalized hypergeometric function, which is defined 
as \cite{HTF}
\bea
_pF_{p-1}(a_1,\cdots,a_p;b_1,\cdots,b_{p-1};z)=
\sum_{n=0}^{\infty}{(a_1)_n\cdots (a_p)_n\over (b_1)_n\cdots (b_{p-1})_n}
{z^n\over n!},\ \ (a)_n={\Gamma(a+n)\over \Gamma(a)}.
\ena
Integrating with respect to 
$\ti u$ of (2.5) we have Higgs field $a$ up to mass residue 
in the weak coupling region in the following form
\bea
a&=&{\sqrt 3 \ti u^{1/2}\over \pi}\sum_{m=0}^
{\infty}{\Gamma(m+{1\over 2})\Gamma({1\over 3})\Gamma({2\over 3})\over
\Gamma({m+1\over 2})\Gamma({m\over 2}+1)m!}
\left({\La\ti m\over 8 \ti u}\right)^m\\
& &\hspace{4cm}\times \, _3F_2\left({1\over 3},
{2\over 3},m-{1\over 2};{m+1\over 2}
,{m\over 2}+1;-{27\La^2\over 16\ti u}\right).\nonumber
\ena
The analytic continuation from this expression 
to around the conformal point 
can be performed to obtain 
\bea
a&=&\sqrt{\pi}\ti u^{1/2}\left({27\La^2\over 16\ti u}\right)^{-1/3}
\sum_{m=0}^{\infty}\left\{{\Gamma({1\over 3})\Gamma({1\over 3})\Gamma(m-
{5\over 6})\over \Gamma({11\over 6})\Gamma(-{1\over 6})\Gamma(m+{1\over 3})m!}
\left({\La \ti m\over 4\ti u}\right)^m\right. \nonumber \\
& & \hspace{5cm}\times 
\  _3F_2\left({1\over 3},-{m\over 2}+{5\over 6},-{m\over 2}+{1\over 3};
{2\over 3},-m+{11\over 6};
-{16\ti u\over 27\La^2}\right) \nonumber\\
& & - \left({27\La^2\over 16\ti u}\right)^{-1/3}
{\Gamma({2\over 3})\Gamma(-{1\over 3})\Gamma(m-{7\over 6})\over 
\Gamma(-{7\over 6})\Gamma({13\over 6})\Gamma(m-{1\over 3})m!}
\left({\La \ti m\over 4\ti u}\right)^m \\
& &\hspace{4cm} \left. \times\ 
_3F_2\left({2\over 3},-{m\over 2}+{7\over 6},-{m\over 2}+{2\over 3};
{4\over 3},-m+{13\over 6};
-{16\ti u\over 27\La^2}\right)
 \right\}. \nonumber 
\ena
If we set $\ti m=0$ we can recover the previous result \cite{MS1} where 
$a$ is represented by the  generalized hypergeometric function $_3F_2$ 
in terms of $\ti u$. 

Next we consider $a_D$. In this case we integrate (2.4) from 
$z=0$ to $z=1$ and evaluate double poles which give the logarithmic 
terms \cite{MS2}. Quite similarly $a_D$ in the weak coupling region 
can be written as
\bea
{\pa a_D\over \pa u}&=&{-1\over 2(-)^{1\over 2}
\ti u^{1\over 2}\pi}\sum_{n,m}
{\Gamma(n+m+{1\over 2})\Gamma(3n+1)\over \Gamma({1\over 2})
 \Gamma(n+1)^2\Gamma(2n+m+1)}\left({\La^2\over 4\ti u}\right)^n
\left(\La\ti m\over 4\ti u\right)^m\nonumber \\
& &\times\left[\psi(n+m+{1\over 2})+3\psi(3n+1)-2\psi(n+1)-
2\psi(2n+m+1)+\ln\left(\La^2\over 4\ti u\right)\right],
\ena
where $\psi(z)$ is defined by
${d\Gamma(z)\over dz}=\psi(z)\Gamma(z)$. 
Analytic continuation to the region $\ti u\sim 0$ and 
integration with respect to $\ti u$ give the 
expression up to mass residue around the conformal point as follows
\bea
a_D&=&{-\sqrt \pi\ti u^{1\over 2}\over (-1)^{1\over 2}}
\left({27\La^2\over 16\ti u}\right)^{-1/3}
\sum_{m=0}^{\infty}\left\{{\Gamma({1\over 3})\Gamma({1\over 3})\Gamma(m-
{5\over 6})\over \Gamma({11\over 6})\Gamma(-{1\over 6})\Gamma(m+{1\over 3})m!}
\left({\La m\over 4\ti u}\right)^m\right. \nonumber \\
& & \hspace{5cm}\times
\  _3F_2\left({1\over 3},-{m\over 2}+{5\over 6},-{m\over 2}+{1\over 3};
{2\over 3},-m+{11\over 6};
-{16\ti u\over 27\La^2}\right) \nonumber\\
& &  +\left({27\La^2\over 16\ti u}\right)^{-1/3}
{\Gamma({2\over 3})\Gamma(-{1\over 3})\Gamma(m-{7\over 6})\over
\Gamma(-{7\over 6})\Gamma({13\over 6})\Gamma(m-{1\over 3})m!}
\left({\La m\over 4\ti u}\right)^m \\
& &\hspace{4cm} \left. \times\
_3F_2\left({2\over 3},-{m\over 2}+{7\over 6},-{m\over 2}+{2\over 3};
{4\over 3},-m+{13\over 6};
-{16\ti u\over 27\La^2}\right)
 \right\}. \nonumber
\ena
As the parameter approaching the point ${\ti m\over \ti u}\rightarrow 
0,\ {\ti u\over \La^2}\rightarrow 0$, we find $a=a_D$, which 
implies that the theory is completely free theory. 
 Therefore the conformal point is certainly 
the fixed point where the beta function of the theory vanishes. 
Since $a,\ a_D\sim \ti u^{5\over 6}$ near the conformal point 
and $a,\ a_D$ are propotional to mass scale of the theory, 
the conformal dimention of $\ti u$ is ${6\over 5}$, which 
has been observed in \cite{APSW}. 

In the $N_f=2$ theory, we use the curve of forth order:
\bea
y^2&=&(x^2-u+{\La^2\over 8})^2-\La^2(x+m_1)(x+m_2)\\
&=&(x^2-u+{\La^2\over 8})^2-\La^2(x^2+Mx+N),
\ena
where we introduce symmetrized mass parameters $M=m_1+m_2$ and 
$N=m_1m_2$. 
 We shift the parameters from the conformal point as
$\ti u=u-{3\La^2\over 8},\ \ \ti M=M-\La,\ \ 
\ti N=N-{\La^2\over 4}$, and rescale $x+{\La\over 2}=\ti u^{1/2}z$,
 we find that the curve can be written  as 
\bea
y^2=\ti u^2(z^2-1)^2-2\La\ti u^{3\over 2}(z^3-z)-
\La^2\ti M\ti u^{1\over 2}z-\La^2\ti N',
\ena
where $\ti N'=\ti N-\ti M\La/2$. 
 As is the case of $N_f=1$, we evaluate the period in the 
weak coupling region by expanding with respect to 
 $1/\ti u$ and mass parameters, and by picking up poles at $z=1$ along 
the $\alpha$ cycle to find
\bea
{\pa a\over \pa u}&=&
{\ti u^{-{1\over 2}}\over 2\pi}
\sum_{m,l=0}^{\infty}{\Gamma(2\alpha_{l,m}+{1\over 2})
\Gamma(\beta_{l,m}+{1\over 2})
\over \Gamma(l+1)\Gamma(m+1)\Gamma(4\alpha_{l,m}+1)}
\left(-{\La^4\ti M^2\over 4\ti u^3}\right)^l
\left({\La^2\ti N'\over \ti u^2}\right)^m
 \\
& & \hspace{1cm} \times
_4F_3\left(\alpha_{l,m}+{1\over 4},\alpha_{l,m}+{3\over 4},
l+{1\over 2},\beta_{l,m}+{1\over 2};
{1\over 2},2\alpha_{l,m}+{1\over 2},2\alpha_{l,m}+1
;-{\La^2\over \ti u}\right) \nonumber \\
&+&{\ti u^{-{1\over 2}}\over 2\pi}\left({2\La\over \ti u^{1\over 2}}
\right)\left({\La^2\ti M\over \ti u^{3\over 2}}\right)
\sum_{m,l=0}^{\infty}{\Gamma(2\alpha_{l,m}+{5\over 2})
\Gamma(\beta_{l,m}+{5\over 2})\over
\Gamma(l+1)\Gamma(m+1)\Gamma(4\alpha_{l,m}+4)}
\left(-{\La^4\ti M^2\over 4\ti u^3}\right)^l
\left({\La^2\ti N'\over \ti u^2}\right)^m
\nonumber \\
& &\hspace{1cm} \times _4F_3\left(\alpha_{l,m}+{5\over 4},\alpha_{l,m}+{
7\over 4},l+{3\over 2},\beta_{l,m}+{5\over 2};{3\over 2},
2\alpha_{l,m}+{5\over 2},2\alpha_{l,m}+2;-{\La^2\over
\ti u}\right),\nonumber
\ena
where $\alpha_{l,m}=l+{m\over 2}$, $\beta_{l,m}=2m+3l$.
  In the $N_f=2$ theory, ${\pa a\over \pa u}$ has
a additional part which is propotional to $\ti M$ and
 vanishes  when  $\ti M=0$.

Next we consider ${\pa a_D\over \pa u}$. 
Performing the line integral from $z=0$ to $z=1$ and evaluating double 
poles, we have ${\pa a_D\over \pa u}$ in the 
weak coupling region as
\bea
{\pa a_D\over \pa u}&=&
{\ti u^{-{1\over 2}}\over 4\pi^2 i}
\sum_{l,m,n}{\Gamma(2\alpha_{l,m}+{1\over 2})
\Gamma(\beta_{l,m}+{1\over 2})
\over \Gamma(l+1)\Gamma(m+1)\Gamma(4\alpha_{l,m}+1)}
\left(-{\La^4\ti M^2\over 4\ti u^3}\right)^l
\left({\La^2\ti N'\over \ti u^2}\right)^m 
\left(-{\La^2\over \ti u}\right)^n\nonumber \\
& &\hspace{1cm}
\times {(\alpha_{l,m}+{1\over 4})_n(\alpha_{l,m}+{3\over 4})_n(l+{1\over 2})_n
(\beta_{l,m}+{1\over 2})_n\over ({1\over 2})_n(2\alpha_{l,m}+
{1\over 2})_n(2\alpha_{l,m}+1)_n}
\left[
\ln\left(-{\La^2\over \ti u}\right)\right.\\
& &\hspace{2cm}+\psi_n(\alpha_{l,m}+{1\over 4})+\psi_n(\alpha_{l,m}+{3\over 4})
+\psi_n(\beta_{l,m}+{1\over 2})\nonumber \\
& &\hspace{3cm}\left.+\psi_n(l+{1\over 2}) -\psi_n(1)-\psi_n({1\over 2})-
\psi_n(2\alpha_{l,m}+{1\over 2})-
\psi_n(2\alpha_{l,m}+1)\right]\nonumber \\
&+&{\ti u^{-{1\over 2}}\over 4\pi^2 i}\left({2\La\over \ti u}\right)
\left({\La^2\ti M\over \ti u^{3\over 2}}\right)
\sum_{l,m,n}{\Gamma(2\alpha_{l,m}+{5\over 2})
\Gamma(\beta_{l,m}+{5\over 2})\over
\Gamma(l+1)\Gamma(m+1)\Gamma(4\alpha_{l,m}+4)}
\left(-{\La^4\ti M^2\over 4\ti u^3}\right)^l
\left({\La^2\ti N'\over \ti u^2}\right)^m
\nonumber \\
& &\hspace{1cm}\times\left(-{\La^2\over \ti u}\right)^n 
{(\alpha_{l,m}+{5\over 4})_n(\alpha_{l,m}+{7\over 4})_n(l+{3\over 2})_n
(\beta_{l,m}+{5\over 2})_n\over 
({3\over 2})_n(2\alpha_{l,m}+{5\over 2})_n(2\alpha_{l,m}+2)_n}
\left[
\ln\left(-{\La^2\over \ti u}\right)\right. \nonumber \\
& &\hspace{2cm}+\psi_n(\alpha_{l,m}+{5\over 4})+
\psi_n(\alpha_{l,m}+{7\over 4})+
\psi_n(\beta_{l,m}+{5\over 2})\nonumber \\
& &\hspace{3cm}\left.+\psi_n(l+{3\over 2})-\psi_n(1)-\psi_n({3\over 2})-
\psi_n(2\alpha_{l,m}+2)-\psi_n(2\alpha_{l,m}+{5\over 2})\right],\nonumber
\ena
where $\psi_n(\alpha)=\psi(n+\alpha)$. 

Analytic continuation of ${\pa a\over \pa u}$ and ${\pa a_D
\over \pa u}$ to the region $\ti u\sim 0$ gives four kinds of $_4F_3$, 
and $a$ and $a_D$ are also represented by $_4F_3$ after 
integration with respect to $\ti u$. 
By defining $\Phi$ as 
\bea
\Phi(\delta,\e;\rho,\sigma,\mu)&=&\,_4F_3
\left(-\alpha_{l,m}+\delta,-\alpha_{l,m}+\delta+{1\over 2},
\alpha_{l,m}+\e,\alpha_{l,m}+\e+{1\over 2}\right. \\
& &\hspace{5cm}\left.;\alpha_{l,m}-\beta_{l,m}+\rho,{m\over 2}+\sigma,\mu;
-{\ti u\over \La^2}\right),\nonumber
\ena
and using this function, 
we find that  $a$ 
around the conformal point can be written in the form: 
\bea
a&=&{\sqrt \pi\ti u^{1\over 2}\over 2}
\sum_{m,l}\left(-{\La^2\ti M^2\over \ti u^2}\right)^l
\left({\La\ti N'\over 2\ti u^{3\over 2}}\right)^m
{1\over \Gamma({1\over 2})\Gamma(m+1)\Gamma(2l+1)\sqrt 2}
\nonumber \\
&\times &\left[c_1
\left({\La^2\over \ti u}\right)^{-{1\over 4}}
{\Gamma(-{m\over 2}+{1\over 4})\Gamma(\beta_{l,m}-\alpha_{l,m}+{1\over 4})
\over \Gamma(\alpha_{l,m}+{3\over 4})
\Gamma({1\over 4}-\alpha_{l,m})}\right.
\Phi\left({1\over 4},{1\over 4};{7\over 4},{3\over 4},{1\over 2}\right)
\nonumber \\
& &-c_2\left({\La^2\over \ti u}\right)^{-{3\over 4}}
{\Gamma(2\alpha_{l,m}+{1\over 2})\Gamma(-{m\over 2}-{1\over 4})
\Gamma(\beta_{l,m}-\alpha_{l,m}-{1\over 4})\over
\Gamma(\alpha_{l,m}+{1\over 4})
\Gamma(-\alpha_{l,m}-{1\over 4})\Gamma(2\alpha_{l,m}-{1\over 2})}
\Phi\left({3\over 4},{3\over 4};{9\over 4},{5\over 4},{3\over 2}\right)
 \\
& &+c_3\left({\La^2\over \ti u}\right)^{-{3\over 4}}
\left({\La^2\ti M\over 2\ti u^{3\over 2}}
\right)
{\Gamma(2\alpha_{l,m}+{5\over 2})\Gamma({1\over 4}-{m\over 2})
\Gamma(\beta_{l,m}-\alpha_{l,m}+{5\over 4})\over
\Gamma(\alpha_{l,m}+{3\over 4})\Gamma({1\over 4}-\alpha_{l,m})
\Gamma(2\alpha_{l,m}+{3\over 2})}
\Phi\left(-{1\over 4},{3\over 4};{3\over 4},{3\over 4},
{1\over 2}\right)
\nonumber\\
& &\left.-c_4\left({\La^2\over \ti u}\right)^{-{5\over 4}}
\left({\La^2\ti M\over 2\ti u^{3\over 2}}
\right)
{\Gamma(2\alpha_{l,m}+{5\over 2})\Gamma(-{m\over 2}-{1\over 4})
\Gamma(\beta_{l,m}-\alpha_{l,m}+{3\over 4})
\over
\Gamma(\alpha_{l,m}+{1\over 4})\Gamma(-\alpha_{l,m}-{1\over 4})
\Gamma(2\alpha_{l,m}+{3\over 2})}
\Phi\left({1\over 4},{5\over 4};{5\over 4},{5\over 4},{3\over 2}\right)\right]
,\nonumber
\ena
where $c_1=c_2=c_3=c_4=1$. 
 We find that the expression for $a_D$ 
is given by changing $c_i$ as $c_1=c_3=(-1)^m,\  c_2=c_4=-(-1)^m$. 
If we set $\ti M=\ti N'=0$ we can recover the previous result \cite{MS1}. 
As in the  $N_f=1$ theory, 
we see that the conformal point is the fixed point 
of this theory from  the relation $a\sim a_D$ on this point. 
Reading the leading power of the expression (2.17), we see that 
the conformal dimension of $\ti u$ is ${4\over 3}$ \cite{APSW}.

In the $N_f=3$ theory, the curve is given by
\bea
y^2&=&(x^2-u+{\La\over 4}x+{(m_1+m_2+m_3)\La\over 8})^2
-\La(x+m_1)(x+m_2)(x+m_3) \nonumber \\
&=&(x^2-u+{\La\over 4}x+{\La L\over 8})^2-\La(
x^3+Lx^2+Mx+N),
\ena
where $L=m_1+m_2+m_3,\ \ M=m_1m_2+m_2m_3+m_3m_1,\ \ 
N=m_1m_2m_3$.
 We shift the parameter from the conformal point as 
$u'=u-{\La^2\over 32},\ \ \ti L=L-{3\La\over 8},\ \ 
\ti M=M-{3\La^2\over 64},\ \ 
\ti N=N-{\La^3\over 512}$, the curve becomes 
\bea
y^2=(x+{\La\over 8})^3(x-{7\La\over 8})-2(u'-
{\La\ti L\over 8})(x+{\La\over 8})^2-
\La(\ti Lx^2+\ti Mx+\ti N) +(u'-{\La\ti L\over 8})^2.
\ena
Setting $\ti u=u'-{\La\ti L\over 8}$ and rescaling  
$x+{\La\over 8}=\ti u^{1/2}z$, we find that the curve can be 
written as 
\bea
y^2=\ti u^2(z^2-1)^2-\ti u^{3\over 2}\La z^3-
\ti L\La \ti u z^2-\ti u^{1\over 2} Az+
B,
\ena
where $A=\La\ti M-{\La^2\ti L\over 4},\ B=
{\La^2\ti L^2\over 64}+{\La^2\ti M\over 8}
-\La\ti N$.
 Evaluation of the integral for the period and 
analytic continuation from $\ti u\sim \infty$ to $\ti u\sim 0$ are same as 
 $N_f=2$ case. In this way,  we can obtain the period in the weak coupling 
region in the form: 
\bea
{\pa a\over \pa u}&=& {\ti u^{-{1\over 2}}\over 2\sqrt \pi}
\sum_{l,m,p,q}^{\infty}
{\Gamma(3\eta_{l,p}+{1\over 2})
\Gamma(\omega_{l,p,q}+{1\over 2})\over 
\Gamma({1\over 2})\Gamma(2\chi_{l,p,q}+1)
l!m!(2p)!}\left({\La\ti L\over \ti u}\right)^l
\left({ A^2\over \ti u^3}\right)^p
\left({B\over \ti u^2}\right)^q  
\nonumber \\
& &  \times \  _4F_3\left(\eta_{l,p}+{1\over 6},
\eta_{l,p}+{1\over 2},\eta_{l,p}
+{5\over 6}, \omega_{l,p,q}+{1\over 2}\right. \\
& &\hspace{5cm}\left. ;{1\over 2},\chi_{l,p,q}+{1\over 2},\chi_{l,p,q}+1;
-{27\La^2\over 256\ti u}\right).\nonumber \\
&-& {\ti u^{-{1\over 2}}\over 2\sqrt \pi}
\left({\La A\over
\ti u^2}\right)
\sum_{l,m,p,q}^{\infty}
{\Gamma(3\eta_{l,p}+{5\over 2})
\Gamma(\omega_{l,p,q}+{5\over 2})\over
\Gamma({1\over 2})\Gamma(2\chi_{l,p,q}+3)
l!m!(2p+1)!}\left({\La\ti L\over \ti u}\right)^l
\left({ A^2\over \ti u^3}\right)^p
\left({B\over \ti u^2}\right)^q
\nonumber \\
& & \times \   _4F_3\left(\eta_{l,p}+{5\over 6},
\eta_{l,p}+{7\over 6},\eta_{l,p}
+{9\over 6}, \omega_{l,p,q}+{5\over 2}\right. \\
& &\hspace{5cm}\left. ;{3\over 2},\chi_{l,p,q}+{3\over 2},\chi_{l,p,q}+2;
-{27\La^2\over 256\ti u}\right),\nonumber
\ena
\bea
{\pa a_D\over \pa u}&=& {\ti u^{-{1\over 2}}\over 4\pi^2 i}
\sum_{l,m,p,q}^{\infty}
{\Gamma(3\eta_{l,p}+{1\over 2})
\Gamma(\omega_{l,p,q}+{1\over 2})\over
\Gamma(2\chi_{l,p,q}+1)
l!m!(2p)!}\left({\La\ti L\over \ti u}\right)^l
\left({ A^2\over \ti u}\right)^p
\left({B\over \ti u^2}\right)^q
\nonumber \\
& &  \times \sum_{n=0}^{\infty}
{(\eta_{l,p}+{1\over 6})_n(\eta_{l,p}+{1\over 2})_n (
\eta_{l,p}+{5\over 6})_n (\omega_{l,p,q}+{1\over 2})_n\over 
({1\over 2})_n(\chi_{l,p,q}+{1\over 2})_n(\chi_{l,p,q}+1)_n n!}
\left(-{27\La^2\over 256\ti u}\right)^n
\nonumber \\
& & \hspace{1cm}\times 
\left[\ln\left(-{27\La^2\over 256\ti u}\right)+
\psi_n(\eta_{l,p}+{1\over 6})+\psi_n(\eta_{l,p}+{1\over 2}) 
+\psi_n(\eta_{l,p}+{5\over 6})\right. \\
& & \hspace{1cm} \left. +\psi_n(\omega_{l,p,q}+{1\over 2})
-\psi_n({1\over 2})-\psi_n(\chi_{l,p,q}+{1\over 2})-\psi_n(
\chi_{l,p,q}+1)-\psi_n(1)\right]\nonumber \\
&-& {\ti u^{-{1\over 2}}\over 4\pi^2 i}
\left({\La A\over
\ti u^2}\right)
\sum_{l,m,p,q}^{\infty}
{\Gamma(3\eta_{l,p}+{5\over 2})
\Gamma(\omega_{l,p,q}+{5\over 2})\over
\Gamma(2\chi_{l,p,q}+3)
l!m!(2p+1)!}\left({\La\ti L\over \ti u}\right)^l
\left({ A^2\over \ti u}\right)^p
\left({B\over \ti u^2}\right)^q
\nonumber \\
& &\times \sum_{n=0}^{\infty}
{(\eta_{l,p}+{5\over 6})_n(\eta_{l,p}+{7\over 6})_n (
\eta_{l,p}+{9\over 6})_n (\omega_{l,p,q}+{5\over 2})_n\over
({3\over 2})_n(\chi_{l,p,q}+{3\over 2})_n(\chi_{l,p,q}+2)_n n!}
\left(-{27\La^2\over 256\ti u}\right)^n
\nonumber \\
& & \hspace{1cm}\times
\left[\ln\left(-{27\La^2\over 256\ti u}\right)+
\psi_n(\eta_{l,p}+{5\over 6})+\psi_n(\eta_{l,p}+{7\over 6})
+\psi_n(\eta_{l,p}+{9\over 6})\right.\nonumber \\
& & \hspace{1cm} \left. +\psi_n(\omega_{l,p,q}+{5\over 2})
-\psi_n({3\over 2})-\psi_n(\chi_{l,p,q}+{3\over 2})-\psi_n(
\chi_{l,p,q}+2)-\psi_n(1)\right], \nonumber 
\ena
where
\bea
\eta_{l,p}={l\over 3}+{p\over 3},\ \
\omega_{l,p,q}=l+3p+2q,\ \
\chi_{l,p,q}={l\over 2}+p+{q\over 2}.
\ena

By analytic continuation and by integration with respect to $\ti u$, 
we obtain $a$ 
around the conformal point in the form:
\bea
a&=&-{2u^{1\over 2}}\sum_{l,p,q}
{2^{\chi_{l,p,q}+1}
\over l!(2p)!q!3^{\eta_{l,p}}}
\left({\La\ti L\over 
\ti u}\right)^l\left({A^2\over \ti u^3}\right)^p
\left({B\over \ti u^2}\right)^q  
\left(-{256\ti u\over 27\La^2}\right)^{-\eta_{l,p}}
 \nonumber \\
&\times &
\left\{{c_1\Gamma({1\over 3})\Gamma({2\over 3})\Gamma(\omega_{l,p,q}-\eta_{l,p}
+{1\over 3})\Gamma(\eta_{l,p}+{1\over 6})
\over \Gamma({1\over 3}-\eta_{l,p})
\Gamma({1\over 3}+\chi_{l,p,q}-\eta_{l,p})\Gamma(
{5\over 6}+\chi_{l,p,q}-\eta_{l,p})}\left(-{256\ti u\over 27\La^2}\right)^
{1\over 6}\right.
\Psi\left({1\over 6},{1\over 6};{1\over 3},{2\over 3},{5\over 3}\right)
\nonumber \\
& &+{c_2\Gamma(-{1\over 3})\Gamma({1\over 3})\Gamma(\omega_{l,p,q}-
\eta_{l,p})\Gamma(\eta_{l,p}+{1\over 2})\over 
\Gamma(-\eta_{l,p})\Gamma(\chi_{l,p,q}-\eta_{l,p})
\Gamma({1\over 2}+\chi_{l,p,q}-\eta_{l,p})}
\left(-{256\ti u\over 27\La^2}\right)^{1\over 2}
\Psi\left({1\over 2},{1\over 2};{2\over 3},{4\over 3},2\right)\\
& &+\left.
{c_3\Gamma(-{2\over 3})\Gamma(-{1\over 3})\Gamma(\omega_{l,p,q}-\eta_{l,p}
-{1\over 3})\Gamma(\eta_{l,p}+{5\over 6})\over 
\Gamma(-\eta_{l,p}-{1\over 3})\Gamma(\chi_{l,p,q}-\eta_{l,p}-{1\over 3})
\Gamma(\chi_{l,p,q}-\eta_{l,p}+{1\over 6})}
\left(-{256\ti u\over 27\La^2}\right)^{5\over 6}
\Psi\left({5\over 6},{5\over 6}
;{4\over 3},{5\over 3},{7\over 3}\right)\right\}
\nonumber\\
&+&{2\pi\sqrt{\pi}\La A\over
\ti u^{1\over 2}}\sum_{l,p,q}
{2^{\chi_{l,p,q}+2}
\over l!(2p)!q!3^{\eta_{l,p}+2}}
\left({\La\ti L\over
\ti u}\right)^l\left({A^2\over \ti u^3}\right)^p
\left({B\over \ti u^2}\right)^q 
\left(-{256\ti u\over 27\La^2}\right)^{-\eta_{l,p}}
\nonumber \\
&\times &
\left\{{c_4\Gamma({1\over 3})\Gamma({2\over 3})\Gamma(\omega_{l,p,q}-\eta_{l,p}
+{5\over 3})\Gamma(\eta_{l,p}+{5\over 6})
\over \Gamma({2\over 3}-\eta_{l,p})
\Gamma({2\over 3}+\chi_{l,p,q}-\eta_{l,p})\Gamma(
{7\over 6}+\chi_{l,p,q}-\eta_{l,p})}\left(-{256\ti u\over 27\La^2}\right)^
{5\over 6}\right.
\Psi\left({1\over 3},-{1\over 6};{1\over 3},{2\over 3},{1\over 3}\right)
\nonumber \\
& &+{c_5\Gamma(-{1\over 3})\Gamma({1\over 3})\Gamma(\omega_{l,p,q}-
\eta_{l,p}+{4\over 3})\Gamma(\eta_{l,p}+{7\over 6})\over
\Gamma({1\over 3}-\eta_{l,p})\Gamma({1\over 3}+\chi_{l,p,q}-\eta_{l,p})
\Gamma({5\over 6}+\chi_{l,p,q}-\eta_{l,p})}
\left(-{256\ti u\over 27\La^2}\right)^{7\over 6}
\Psi\left({2\over 3},{1\over 6};{2\over 3},{4\over 3},{2\over 3}\right)
\nonumber \\
& &+\left.
{c_6\Gamma(-{2\over 3})\Gamma(-{1\over 3})\Gamma(\omega_{l,p,q}-\eta_{l,p}
+1)\Gamma(\eta_{l,p}+{3\over 2})\over
\Gamma(-\eta_{l,p})\Gamma(\chi_{l,p,q}-\eta_{l,p})
\Gamma(\chi_{l,p,q}-\eta_{l,p}+{1\over 2})}
\left(-{256\ti u\over 27\La^2}\right)^{3\over 2}
\Psi\left(1,1;{4\over 3},{5\over 3},1\right)\right\},\nonumber 
\ena
where 
\bea
\Psi(a,b;c,d,e)=\! _4F_3\left(\eta_{l,p}+a,\eta_{l,p}+a+{1\over 2},
\eta_{l,p}-\chi_{l,p,q}+b\right.
&,&\eta_{l,p}-\chi_{l,p,q}+b+{1\over 2} \\
& &\left. ;
c,d,\eta_{l,p}-\omega_{l,p,q}+e;-{256\ti u\over 27\La^2}\right),\nonumber
\ena
and $c_1=\cdots =c_6=1$. The expression for $a_D$ 
is given by changing $c_i$ as $c_1=\cot(\eta_{l,p}+{1\over 6})\pi,\ 
c_2=\cot(\eta_{l,p}+{1\over 2})\pi,\ c_3=c_4=
\cot(\eta_{l,p}+{5\over 6})\pi,\ c_5=
\cot(\eta_{l,p}+{7\over 6})\pi,\ c_6=\cot(\eta_{l,p}+{3\over 2})\pi$. 
 If we set $\ti L=\ti M=\ti N=0$, i.e., $A=B=0$ and $\ti u=u'=u-
{\La^2\over 32}$, we can recover the previous result \cite{MS1}. 
As was the case of $N_f=1,2$, the relation $a\sim a_D$ hold 
on the conformal point, therefore we can recognize the 
conformal point is the fixed point of this theory.

Let us compare our expressions to the ones 
obtained by the expansion around the different point from 
the conformal point. 
If we consider the massive theory as the generalization 
from the massless theory, 
we would treat the bare mass parameter as the deviation from the 
massless theory. 
In order to see the behaviour of the field $a$ and $a_D$ 
in the weak coupling region in this case, 
we expand the meromorphic differential 
$\la$ with respect to $\La$ and mass parameters, and 
evaluate the integral representation along 
the corresponding cycle. 
For example in the case of $N_f=1$, $\la$, $a$ and $a_D$ are given by 
\bea
\la&=&{x\ dx\over 2\pi i\ y}\left({(x^2-u)\over 2(x+m)}-(x^2-u)'\right),
\nonumber \\
a&=&\oint_{\alpha}\la,\ \ \ \ \ 
a_D=\oint_{\beta}\la,
\ena
where we use the curve of forth order. 
The result of the calculation 
for the field $a$ can be 
written as
\bea
a&=&{\sqrt{u}\over 12 \sqrt{\pi}}\sum_{n,l=0}
{\Gamma(n-{1\over 6})\Gamma(n+{1\over 6})\Gamma(
l-n)\Gamma(l+3n-{1\over 2})\over 
\Gamma(n+1)\Gamma(-n)\Gamma(3n-{1\over 2})\Gamma(l+{1\over 2})n!l!}
\left({m^2\over u}\right)^l\left({
-27\La^6\over 256u^3}\right)^n \\
&+&{3\sqrt u\over 32\sqrt{\pi}}\left({\La^3m\over u^2}\right)
\sum_{n,l=0}{\Gamma(n+{7\over 6})\Gamma(n+{5\over 6})
\Gamma(l-n)\Gamma(l+3n+{3\over 2})\over 
(2n+1)\Gamma(n+1)\Gamma(-n)\Gamma(3n+{3\over 2})\Gamma(l+{3\over 2})
n!l!}\left({m^2\over u}\right)^l\left({
-27\La^6\over 256u^3}\right)^n.\nonumber 
\ena
In the massless limit, this expression 
 reduces to the previous result obtained by 
solving the Picard-Fucks equation \cite{IY}, which 
is represented by using the Gauss' hypergeometric function. 
 The expression (2.28) can be verified by 
expanding the following expression which is represented 
by using the modular invariant form \cite{MS1}:
\bea
{\pa a\over \pa u}&=&(-D)^{-{1\over 4}}F\left({1\over 12},
{5\over 12};1;-{27\Delta\over 4D^3}\right),\nonumber \\
\Delta&=&-\La^6(256u^3-256u^2m^2-288um\La^3+256m^3\La^3+27\La^6),\\
D&=&-16u^2+12m\La^3.\nonumber 
\ena
in the weak coupling region, and by comparing two expressions order by 
order after $u$ integration. 
 In the $N_f=2,3$ case, 
instead of integrating $\la$ to obtain fields $a$ and $a_D$, 
we can evaluate ${\pa a\over \pa u}$ and ${\pa a_D\over \pa u}$ by expanding 
 around the massless point in a similar manner. The results are  
expressed in terms of the following arguments: 
\bea 
{1\over 64}\left({\La^2\over u}\right)^2\ (N_f=2),\hspace{1cm}
{1\over 256}\left({\La^2\over u}\right)\ (N_f=3),
\ena
 and appropriate combinations of mass parameters. 
These are identical to the argument of the hypergeometric 
function describing the massless theories \cite{IY}. 
These powers of $\La$ are equivalent to the powers of 
the instanton term of the curve, and vary as the number 
of matters we have introduced. 
 On the contrary, the argument of the 
expression we have derived in this section 
 is simple compared to (2.29), which is the argument based 
of the the deviation from the conformal point, 
 and the form of these deviations 
does not depent on the number of the matters 
 as we have seen in this section.
Thus if we use the 
parametrization from the conformal point, 
the theory can be described by using the  
simple deviation from the conformal point 
even in such case that we discuss the weak coupling behaviour. 
 Furthermore the expression around the massless point in the
$N_f=1,2$ case can be obtained from our expression for the $N_f=3$ case
 by taking suitable double scaling limit
to decouple the irrelevant mass parameters.
These are obvious advantages to observe the behaviour of the theory 
by using the expression around the conformal point.

Before closing this section, 
we discuss the relation between 4-D $SU(2)$ $N=2$ supersymmetric 
QCD and 2-D $N=2$ SCFT, which has been partially analized 
in our previous paper \cite{MS1}. 
Let us review the Landau-Ginzburg description of 
2-D $N=2$ superconformal minimal models with $c=3$ which describe 
the torus. Since 
the theory with central charge $c=3k/k+2$ corresponds to the 
Landau-Ginzburg potential $x^{k+2}$, we have three type of description;
 $(k=1)^3$, $(k=2)^2$ and $(k=1)(k=4)$, as 
\bea
f_1&=&x^3+y^3+z^3,\nonumber \\
f_2&=&x^4+y^4+z^2, \\
f_3&=&x^6+y^3+z^2. \nonumber 
\ena
These are known as the algebraic curve on the (weighted) complex 
projective space $(W)\bf{CP}^2$ with homogeneous coordinates $[x,y,z]$
 describing 
singular torus, and their typical deformation 
in one parameter $\psi$ are following
\bea
\ti E_6\ :\ &f&={x^3\over 3}+{y^3\over 3}+{z^3\over 3}-\psi_6 xyz=0,
\nonumber \\
\ti E_7\ :\ &f&={x^4\over 4}+{y^4\over 4}+{z^2\over 2}-\psi_7 xyz=0,\\
\ti E_8\ :\ &f&={x^6\over 6}+{y^3\over 3}+{
z^2\over 2}-\psi_8 xyz=0,\ 
\nonumber
\ena
where we have used appropriate normalization.   
 We can evaluate the period ${\cal W}$: 
\bea
{\cal W}=\psi\int_{\Gamma} {dxdydz\over f},
\ena
on each curve in the region $\psi\sim \infty$ by picking up  poles 
of the integrant expanded by $1/\psi$. 
Altanative approach to obtain the period is solving 
the Picard-Fuchs equation corresponding to these curves 
\bea
(1-y)y{d^2{\cal W}\over dy^2}+(1-2y){d{\cal W}\over dy}-{1\over \alpha}
(1-{1\over \alpha})
{\cal W}=0,
\ena
where $y=\psi^{-\alpha}$ and $\alpha=3\ (\ti E_6),\ 
4\ (\ti E_7),\ 6\ (\ti E_8)$. 
As a result, periods are expressed as  
linear combinations of $F({1\over \alpha},1-{1\over \alpha},1;
y)$ and $F^*({1\over \alpha},1-{1\over \alpha},1;
y)$ around $y=0$ where $F$ is Gauss' hypergeometric function $_2F_1$, 
and $F^*$ is another independent solution corresponding to $F$. 
Comparing these results to the expression 
 obtained by setting mass 
deviations zero in the results we have derived in 
this section, or 
the more obvious expression in \cite{MS1}, we find that 
periods of $\ti E_6,\ \ti E_7$ and $\ti E_8$ curves are 
identical to the periods 
${\pa a\over \pa u},\ {\pa a_D\over \pa u}$ of 
4-D $N=2$ supersymmetric $SU(2)$ QCD with $N_f=1,\ 2$ and $3$ 
matter fields respectively in the
weak coupling region $\ti u\sim \infty$ 
when the theory has the conformal point. 
In this way, 
 we can find a simple identification between the moduli parameter of each
theory, which is  
\bea
\psi^{\alpha}\  \longleftrightarrow\ti u,
\ena
up to irrelevant constant factors,
  and Landau-Ginzburg point $\psi=0$ of torus corresponds to 
the fixed point $\ti u=0$ of N=2 supersymmetric SU(2) QCD. 
 This is another confirmation of our expression around the critical points. 
 It is also interesting that another toric description of torus:
\bea
{1\over 2}x^2+{1\over 2}y^2-\psi zw=0,\ \ \ 
{1\over 2}z^2+{1\over 2}w^2-xy=0,
\ena
can be regarded as the curve which corresponds 
to $N_f=0$ curve whose parameter $\psi^2$ 
can be identified by deviation from the dyon point.

\sect{$SU(N)$ pure Yang-Mills theories}

In this section we study pure $SU(N)$ theory. 
 In this case the curve is given by
\bea
y^2=(x^N-\sum_{i=2}^Ns_ix^{N-i})^2-\La^{2N},
\ena
where $s_i\ (2\le i\le N)$ 
are gauge invariant moduli parameter. We 
treat meromorphic differential $\la$ directly,
 and calculate the period of meromorphic
differential $\la$, i.e. Higgs field and its dual, which are  defined by  
\bea
\la={x\over 2\pi i}(x^N-\sum_{i=2}^Ns_ix^{N-i})'{dx\over y}\\
a_i=\oint_{\alpha_i}\la,\ \ \ \  
a_D^i=\oint_{\beta_i}\la.
\ena
We consider so called $Z_N$ critical point $s_2=\cdots=s_{N-1}=0,\ 
s_N=\pm \La^{N}$ where the curve becomes \cite{AD,APSW,EHIY}
\bea
y^2=x^{N}(x^N\pm 2\La^N).
\ena
First we evaluate the integral 
in the region $s_i\sim 0\ (2\le i\le N-1),\ s_N\sim \infty$. To this end, 
we expand the meromorphic differential $\la$ 
with respect to $\La^{2N}$ in the form
\bea
\la={dx\over 2\pi i}\sum_{n=0}^{\infty}{\Gamma(n+{1\over 2})
(\La^{2N})^n\over \Gamma({1\over 2})\ n!\ 2n}
(x^N-\sum_{i=2}^Ns_ix^{N-i})^{-2n}.
\ena
Rescaling $x=s_N^{1/N}z$ and $\alpha_i=s_is_N^{-i/N}$, 
and expanding with respect to $1/s_N$ and $\alpha_i$, 
$\la$ becomes 
\bea
\la&=&{s_N^{1/N}dz \over 2\pi i}\sum_{n=0}^{\infty}
{\Gamma(n+{1\over 2})\over \Gamma({1\over 2})n!2n}\left(
{\La^{2N}\over s_N^2}\right)^n(z^N-1)^{-2n}\nonumber \\
& & \hspace{3cm}
\times \sum_{\{m\}}^{\infty}{\Gamma(a_{\{m\}}+2n)\over \Gamma(2n)}
\prod_{i=2}^{N-1}{1\over m_i !}\left({\alpha_iz^{N-i}\over 
z^N-1}\right)^{m_i},
\ena
where $\{m\}=\{m_2,\cdots,m_{N-1}\}$ and $ a_{\{m\}}=\sum_{i=2}^{N-1}m_i$. 
In order to calculate $a_i$, we pick up the poles at $\dis e^{
{2\pi ik\over N}}$ in meromorphic 
differential along $\alpha_i$ cycle. 
By introducing Barnes-type integral representation \cite{MS2}  
and multiplying $\sin 2s\pi/\pi$,  
we integrate from $z=0$ to $\dis z=e^{{2\pi ik\over N}}$ to pick 
up the poles as
\bea
a_k&=&s_N^{{1\over N}}\int_{-i\infty}^{i\infty}
{ds\over 2\pi i}\sum_{\{m\}}^{\infty}\int_0^{e^{2\pi ik\over N}}
dz {\Gamma(s+{1\over 2})(-1)^s\Gamma(-s)\Gamma(a_{\{m\}}+2s)\over 
\Gamma({1\over 2})2s \Gamma(2s)}{\sin 2s\pi\over \pi}\nonumber\\
& &\hspace{2cm}\times (z^N-1)^{-2s-a_{\{m\}}}\,z^{Na_{\{m\}}-b_{\{m\}}}
\prod_{i=2}^{N-1}
{\alpha_i^{m_i}\over m_i!}\left({\La^{2N}\over s_N^2}\right)^s,
\ena
where $ b_{\{m\}}=\sum_{i=2}^{N-1}i\, m_i$. 
Therefore we find that 
 $a_k$ in the region where $s_N\sim \infty$ is given by
\bea
a_k&=&{s_N^{1\over N}\over N}\sum_{n,\{m_i\}}^{\infty}
{e^{-2\pi ikb'_{\{m\}}}
\Gamma(n+{1\over 2})\Gamma(a_{\{m\}}-b'_{\{m\}})
\over \Gamma({1\over 2})\Gamma(2n+1)n!
\Gamma(-2n-b'_{\{m\}}+1)}
\prod_{i=2}^{N-1}{\alpha_i^{m_i}\over m_i!}\left(
{\La^{2N}\over s_N^2}\right)^n,
\ena
where $b'_{\{m\}}=(b_{\{m\}}-1)/N$. Note that
the phase factor guarantees the constraint $\sum_{i=1}^N a_i=0$.
 In order to continuate analytically 
to the region $s_N\sim \La^{N}$ and to use various identities, we re-express
 (3.8) by using the hypergeometric function as 
\bea
a_k&=&{s_N^{1\over N}\over N}
\sum_{n,\{m_i\}}^{\infty}
{e^{-2\pi ikb'_{\{m\}}}
\Gamma(a_{\{m\}}-b'_{\{m\}})\over \Gamma(1-b'_{\{m\}})}
\prod_{i=2}^{N-1}{\alpha_i^{m_i}\over m_i!}
F\left({b'_{\{m\}}\over 2},{b'_{\{m\}}+1\over 2};1;{\La^{2N}\over s_N^2}
\right).
\ena
Quite generally the expression of $a_k$ differs by the choice of the 
branch, therefore we cannot perform analytic continuation of the 
expression beyond the convergence domain. In the case of $SU(2)$, 
this process can be justified by comparison to the elliptic 
singular curve made for torus. 
 For general hyper-elliptic curve, there is no such 
guarantee for the process. However in our expression, 
${\La^{2N}\over s^2_N}=1$ is the critical point which is just on 
the boundary of the convergence domain, therefore 
we can obtain expression around ${\La^{2N}\over s^2_N}=1$.
 Performing analytic continuation to ${\La^{2N}\over s_N^2}\sim 1$, and 
using the identity for the hypergeometric function
\bea
F(a,b,c,w)=(1-w)^{-a}F(a,c-b,c,{w\over w-1}),
\ena
and the quadratic transformation \cite{HTF}
\bea
F(2a,2b,a+b+{1\over 2},z)=F(a,b,a+b+{1\over 2},4z(1-z)),
\ena
and also using another identity 
\bea
F(a,b,c,z)=(1-z)^{c-a-b}F(c-a,c-b,c,z),
\ena
where $z={1\over 2}-{s_N\over 2\La^N}$, 
we can put  $a_{k}$ around the conformal point in the form:
\bea
a_{k}&=&{s_N^{1\over N}\over N}
\sum_{\{m\}}^{\infty}
{e^{-2\pi ikb'_{\{m\}}}
\Gamma(a_{\{m\}}-b'_{\{m\}})\over \Gamma(1-b'_{\{m\}})}
\prod_{i=2}^{N-1}{\alpha_i^{m_i}\over m_i!}
\left({s_N+\La^N\over 2\La^N}\right)^{1\over 2}\nonumber\\
&\times &\left\{{\Gamma({1\over 2}-b'_{\{m\}})
\over \Gamma(1-b'_{\{m\}})
}\right.
\left({\La^N+s_N\over s_N}\right)^{-b'_{\{m\}}}
F\left({1\over 2},{1\over 2};b'_{\{m\}}+{1\over 2};z\right)
 \label{eq:su(n)ac}\\
& &\left.+{\Gamma(b'_{\{m\}}-{1\over 2})
\over \Gamma(b'_{\{m\}})}
\left({\La^N+s_N\over s_N}\right)^{-{1\over 2}}
\left({s_N-\La^N\over s_N}\right)^{{1\over 2}-b'_{\{m\}}}
F\left({1\over 2},{1\over 2};
{3\over 2}-b'_{\{m\}};z\right)\right\}.
\nonumber
\ena

Next we consider $a_D^i$. In this case we 
integrate from $z=0$ to $z=e^{2\pi i k\over N}$ 
without multiplying $\sin 2s\pi$ as
\bea
a_D^k&=&{s^{1\over N}_N\over \pi iN}
\int_{-i\infty}^{i\infty}{ds\over 2\pi i}
\sum_{\{m\}}e^{-2\pi ikb'_{\{m\}}}
\Gamma(a_{\{m\}}-b'_{\{m\}})(-1)^{-2s-a_{\{m\}}}\\
& &\hspace{1cm}\times{\Gamma(s+{1\over 2})\Gamma(-s)\Gamma(a_{\{m\}}+2s)
\Gamma(-2s-a_{\{m\}}+1)\over \Gamma({1\over 2})\Gamma(2s+1)\Gamma(-2s-
b'_{\{m\}}+1)}
\left(\prod_{i=2}^{N-1}
{\alpha_i^{m_i}\over m_i!}\right)\left(-{\La^{2N}\over s_N^2}\right)^s,
\nonumber 
\ena
which is defined modulo $a_k$ in the weak coupling region. 
We evaluate double poles of this integral and 
also subtract the contribution from $z=0$ \cite{MS2}
to obtain $a_D^k$ 
\bea
a_D^k&=&
{s^{1\over N}_N\over \pi iN}\sum_{\{m\}}^{\infty}{e^{-
2\pi ikb'_{\{m\}}}
\Gamma(a_{\{m\}}-b'_{\{m\}})\over 
\Gamma(1-b'_{\{m\}})}
\left(\prod_{i=2}^{N-1}{\alpha_i^{m_i}\over m_i!}\right)
\nonumber \\
& &\times\sum_{n=0}^{\infty}{({b'_{\{m\}}\over 2})_n
({b'_{\{m\}}\over 2}+{1\over 2})_n\over \Gamma(n+1)n!}
\left({\La^{2N}\over s_N^2}\right)^n \\
& &\times\left\{\ln\left({\La^{2N}\over s_N^2}\right)+
\psi(n+{b'_{\{m\}}\over 2})\right.\nonumber \\
& &\hspace{2cm}\left.+
\psi(n+{b'_{\{m\}}\over 2}+{1\over 2})
-2\psi(n+1)+\pi\cot(b'_{\{m\}}\pi)\right\}.\nonumber
\ena
Using the analytic continuation to the region $s_N\sim \La$, 
we have
\bea
a_D^k&=&
{s^{1\over N}_N\over \pi iN}\sum_{\{m\}}
{e^{-2\pi ikb'_{\{m\}}}
\Gamma(a_{\{m\}}-b'_{\{m\}})\over 
\Gamma(1-b'_{\{m\}})}
\left(\prod_{i=2}^{N-1}{\alpha_i^{m_i}\over m_i!}\right)
\left({s_N+\La^N\over 2\La^N}\right)^{1\over 2}\nonumber \\
& &\times\left\{{\pi\cot(b'_{\{m\}}\pi)
\Gamma(b'_{\{m\}}-{1\over 2})\over 
\Gamma(b'_{\{m\}})
}\right. \label{eq:su(n)adc}\\
& &\hspace{3cm}\times \left.
\left({s_N+\La^N\over s_N}\right)^{-{1\over 2}}
\left({s_N-\La^N\over s_N}\right)^{{1\over 2}-b'_{\{m\}}}
F\left({1\over 2},{1\over 2};{3\over 2}-
b'_{\{m\}};z\right)
\nonumber \right\}.
\ena

Around the critical point, the original roots of the curve 
$e_k^{+},\ e_k^{-}$ which both reduce to $e_k$ for $\La=0$, become 
$ e_k^{+}\simeq \La e^{2\pi ik\over N},\ e_k^{-}\simeq 0$. The 
expression (\ref{eq:su(n)ac}) and (\ref{eq:su(n)adc}) 
show that $a_k$ consists of the contribution from both poles whereas 
$a_D^k$ consists of the contribution from $e_k^{-}$ which vanishes
 at the critical point. 
Of course, we can find an expression for $a_D^k$ which reduces 
to $a_k$ at the conformal point, i.e., $a_D^{'k}=a_D^k+
a_k$,  because $a_D^k$ was defined 
modulo $a_k$, 
which cannot be determined by analytic continuation but by 
the consistency. 
Therefore, around the conformal point $a_k$ and $a_D^k$ behave as
\bea
a_k\sim a_D^k\sim(s_N-\La^N)^{N+2\over 2N}+\  const. 
\ena
From this result, we recognize that the conformal point 
is certainly the fixed point of the theory, and the 
conformal dimension of $s_N$ is ${2N\over N+2}$ \cite{EHIY}. 

We have used the ordinary type of the analytic continuation but the 
presence of the factor $\Gamma(-b'_{\{m\}}+{1\over 2})$ shows that  
this factor has 
 poles and the expression (\ref{eq:su(n)ac}) and (\ref{eq:su(n)adc}) contain
the logarithmic terms. To see this, we decompose 
$b_{\{m\}}$
 mod
 $N$ as $b_{\{m\}}=Nl+\la$ where $l=0,1,2,\cdots$ and $0\le \la\le N-1$. 
Noticing that 
$b'_{\{m\}}=(b_{\{m\}}-1)/N$, when $N$ is even and 
$\la=1+{N\over 2}$, $b'_{\{m\}}-{1\over 2}$ becomes integer, thus we find that 
$\Gamma(-b'_{\{m\}}+{1\over 2})$ has poles. 
That is, around the conformal point of the moduli space of pure $SU(2n)$ 
theory, there are unstable directions that $a_i$ and $a_D^i$ have 
 logarithmic terms. However except these directions $a_i$ and 
$a_D^i$ contain 
no logarithmic term, and since just on the conformal point $\Gamma(
-b'_{\{m\}}+{1\over 2})=\Gamma(-{1\over N}+{1\over 2})$ 
there is no logarithmic singularity except $N=2$, 
 the conformal point is still the fixed point of the theory. 
When we set $N=2$, i.e., gauge group $G=SU(2)$, 
the point we considered is a dyon point. Therefore 
it is natural that $a$ and $a_D$ have such logarithmic contribution. 

As a check of our result and an example, we consider 
the  gauge gourp $G=SU(3)$. 
We set $u=s_2,\ v=s_3,\ \alpha_2=u/v^{2\over 3}$ and $a_{\{m\}}=m,\
b'_{\{m\}}=(2m-1)/3$.
 In the weak coupling region $v\sim \infty$, our expression
reduces to Appell's $F_4$ system \cite{HTF} with argument ${
4u^3\over 27v^2},\ {\La^6\over v^2}$. 
Analytic continuation to the region $u\sim \infty$ 
recovers the result in ref.\cite{KLT} up to 
the choice of branch for the logarithmic term of 
$a_D^i$, which is again  
represented by Appell's $F_4$ system. 
 By analytic continuation to 
around the conformal point, we can find that our expression becomes 
 Horn's $H_7$ system \cite{HTF}. To see this,  
we set $m=3l+\la\ (l=0,1,2,\cdots,\ 
\la=0,1,2)$ so that $a_k$ and $a_D^k$ are  decomposed to $a_k=\sum_{\la=0}^2
a_k^{\la},\ 
a_D^k=\sum_{\la=0}^2a_D^{k\la}$. 
 For example, $a_k^{\la}$ can be written as
\bea
a_k^{\la}&=&{v^{1\over 3}e^{-{2\pi i k\over 3}(2\la-1)}
\sin({2\la-1\over 3})\pi \ 2^{2\la-1\over 3}\over 
i6\pi\Gamma({1\over 2})^3}\left({u^3\over v^2}\right)^{\la\over 3}
\left({v\over \La^3}\right)^{1\over 2}
\sum_{n,l}{\Gamma(l+{\la+1\over 3})
\over \Gamma(3l+\la+1)} \label{eq:su(3)ac}\\
& &\times \left\{ \left({\La^3\over v}\right)^{-{2\la\over 3}+{5\over 6}}
{\Gamma(2l+n+{2\la\over 3}-{1\over 3})^2\sin({2\la\over 3}-{1\over 3})\pi
\  z^n\over 
\Gamma(2l+n+{2\la\over 3}+{1\over 6})
\sin({2\la\over 3}+{1\over 6})\pi\ n!}
\left(u^3\over 4\La^6\right)^l \right.\nonumber \\
& &+\left. \left({2(v-\La^3)\over v}\right)^{-{2\la\over 3}+{5\over 6}}
\Gamma(n+{1\over 2})^2\Gamma(2l-n+{2\la\over 3}-{5\over 6})\ {(-z)^n\over 
n!}\left({u^3\over (v-\La^3)^2}\right)^l\right\},\nonumber
\ena
where $ z={1\over 2}(1-{v\over \La^3})$. 
Because of a factor $\sin ({2\la-1\over 3}) \pi$ in (\ref{eq:su(3)ac}),
 the component for $\la=2$ disappears, i.e., $a_k^2=0$. 
For $\la=0,1$, the second term can be expressed by  Horn's $H_7$ function 
as
\bea
H_7\left(-{5-4\la\over 6},{1\over 2},{1\over 2},
{2+2\la\over 3},{u^3\over 27(v-\La^3)^2},-{1\over 2}(1-{v\over \La^3})\right),
\ena
where $H_7(a,b,c,d,x,y)$ is given by \cite{HTF}
\bea
H_7(a,b,c,d,x,y)=\sum_{n,m}{(a)_{2m-n}(b)_n(c)_n\over 
(d)_m m! n!}x^m y^n.
\ena
This means that if we choose the variable $x={u^3\over 27(v-\La^3)^2}$ 
and $y=-{1\over 2}(1-{v\over \La^3})$, Picard-Fuchs equations of the 
theory should reduce to differential equations 
 of $H_7(a,b,c,d,x,y)$ system which is 
given by \cite{HTF}
\bea
& \left\{-y(1+y)\pa^2_y+2x{\pa_x\pa_y}+(a-1-(b+c+1)y)\pa_ y-bc\right\}H_7=0,
\label{eq:h7eq} \\
&\left\{x(1-4x)\pa^2_x+4xy\pa_x\pa_y-y^2
\pa^2_y+(d-(4d+6)x)\pa_x+2ay\pa_y-a(a+1)\right\}H_7=0,\nonumber
\ena
where we have corrected a misprint in ref.\cite{HTF}. 
Furthermore, noticing that four independent solutions 
of this system can be written as 
\bea
& &H_7(a,b,c,d,x,y)\nonumber \\
&x&^{1-d}H_7(a-2d+2,b,c,2-d,x,y),\nonumber \\
&y&^a\sum_{m,n=0}^{\infty}{(b+a)_{2m+n}(c+a)_{2m+n}\over 
(d)_m(1+a)_{2m+n}m!n!}(xy^2)^m(-y)^n,\label{eq:h7}\\
&y&^{a-2d+2}\sum_{m,n}^{\infty}{(b+a-2d+2)_{2m+n}(c+a-2d+2)_{2m+n}
\over (2-d)_m(a-2d+3)_{2m+n}}(xy^2)^m(-y)^n,\nonumber
\ena
first and second term of (\ref{eq:su(3)ac}) with $\la=0,1$ 
correspond to above solutions 
of this system. 
 Let us chech this point. 
We start with the Picard-Fuchs equation in this theory for 
$\Pi=\oint \la$ \cite{KLT}:
\bea
{\cal L}_1\Pi&=&
\left\{(27\La^6-4u^3-27v^2)\pa_u^2-12u^2v\pa_u\pa_v-3uv\pa_v-u\right\}
\Pi=0,\nonumber \\
{\cal L}_2\Pi&=&
\left\{(27\La^6-4u^3-27v^2)\pa_v^2-36uv\pa_u\pa_v-9v\pa_v-3\right\}
\Pi=0.\label{eq:su(3)pf}
\ena
By direct change of variables 
 $x={u^3\over 27(v-\La^3)^2}$
 and $y=-{1\over 2}(1-{v\over \La^3})$, and some linear combinations of 
these equations, 
we can check that the Picard-Fuchs equation (\ref{eq:su(3)pf}) can 
be written as 
\bea
&x&(1-4x)\pa^2_x\Pi_0-y^2\pa^2_y\Pi_0+4xy\pa_x\pa_y\Pi_0+{2\over 3}
(1-4x)\pa_x\Pi_0-{5\over 3}y\pa_y\Pi_0+
{5\over 36}\Pi_0=0,\nonumber\\
&y&(1+y)\pa^2_y\Pi_0-2x\pa_x\pa_y\Pi_0+
{11+12y\over 6}\pa_y\Pi_0+{1\over 4}\Pi_0=0.
\ena
where $\Pi_0=y^{-{5\over 6}}\Pi$. Comparing this to 
(\ref{eq:h7eq}), we see that this system is identical to (\ref{eq:h7eq}) with
 $a=-{5\over 6},\ b=c={1\over 2},\ d={2\over 3}$. Substituting 
these to (\ref{eq:h7}), we can find directly that 
four solutions of the Picard-Fuchs equation of this theory 
are identical to four functions of the expression (\ref{eq:su(3)ac}) with 
$\la=0,1$, 
 although the first term of (\ref{eq:su(3)ac}) are not within the Horn's list.

\sect{$SO(2N)$ pure Yang-Mills theories}

In this section we discuss pure $SO(2N)$ theory 
whose singular points in the 
strong coupling region are known in arbitrary 
$N$ \cite{EHIY}. 

In pure $SO(2N)$ theory the curve and meromorphic differential are given by
\bea
y^2&=&P(x)^2-\La^{4(N-1)}x^4=
\left(x^{2N}-\sum_{i=1}^Nx^{2(N-i)}s_i\right)^2-\La^{4(N-1)}x^4,\\ 
\la&=&(2P(x)-xP'(x)){dx\over y}.
\ena
Since the difference from $SU(N)$ theory 
is only powers of $\La$ in the instanton correction term, 
the calculation is almost same as $SU(N)$ theory. 
What we need is the expression around the point $
s_i=0 \, (i\ne N-1),\  s_{N-1}=\pm\La^{2N-2}$ 
where the curve is degenerate as \cite{EHIY}
\bea
y^2=x^{2N+2}(x^{2N-2}\pm 2\La^{2N-2}).
\ena
To this end, it is convenient to evaluate integral in the region 
$s_i\sim 0\, (i\ne N-1),\  s_{N-1}\sim \infty$. 
Expanding $\la$ with respect to $\La^{4(N-1)}$ and integrating 
by part, we can rewrite  $\la$ in the following form:
\bea
\la=\int_{-i\infty}^{\infty}
{ds\over 2\pi i}{dx\over 2\pi i}
{\Gamma(s+{1\over 2})\Gamma(-s)\over 
\Gamma({1\over 2})2s}\left(-\La^{4(N-1)}x^4\right)^sP(x)^{-2s},
\ena
where we have introduced  Barnes-type integral representation as before. 
 Rescaling the variable as 
\bea
x=s_{N-1}^{1/(2N-2)}z=uz, \ \ 
s_i=u^{2i}\alpha_i\, (i\ne N-1).
\ena
and expanding with respect to $\alpha_i$ and $\La^{4N-4}/u^{4N-4}$, 
we have $\la$ in the form: 
\bea
\la&=&u\int_{-i\infty}^{i\infty}{ds\over 2\pi i}{\Gamma(s+{1\over 2})
\Gamma(-s)\Gamma(2s+a_{\{m\}})\over \Gamma({1\over 2})\Gamma(2s+1)}
\left(-{\La^{4N-4}\over 
u^{4N-4}}\right)^s\sum_{\{m\}}\prod_{i\ne N-1}^{N}
\left({\alpha_i\over m_i!}\right)^{m_i}\\
& &\hspace{2cm}
\times \int{dz\over 2\pi i}
z^{2(N-1)a_{\{m\}}-2b_{\{m\}}}(z^{2N-2}-1)^{-2s-a_{\{m\}}},\nonumber
\ena
where $\{m\}=\{m_1,\cdots,m_{N-2},m_N\}$ and $
a_{\{m\}}=\sum_{i=1,i\ne N-1}^{N}m_i,\ \ 
b_{\{m\}}=\sum_{i=1,i\ne N-1}^{N}im_i$. 
In order to obtain $a_k$, we pick up poles at $z=
e^{2\pi i k\over 2N-2}\ 
(0\le k\le N-1)$ along $\alpha_k$ cycle and $z=0$ 
along $\alpha_N$ cycle. 
First we calculate $a_k\  (0\le k\le N-1)$. 
To pick up poles
 at $z=e^{2\pi i k\over 2N-2}$ we integrate 
from $z=0$ to $z=
e^{2\pi i k\over 2N-2}$ multiplying $\sin 2s \pi/\pi$ to find that 
 $a_k$ can be expressed in the form:
\bea
a_k&=&{u\over 2N-2}\sum_{n,\{m\}}^{\infty}{e^{-2\pi ikb'_{\{m\}}}
\Gamma({1\over 2}+n)\over \Gamma({1\over 2})\Gamma(2n+1)
n!}\left({\La^{4N-4}\over u^{4N-4}}\right)^n\prod_{i\ne N-1}
\left({\alpha_i\over m_i!}\right)^{m_i}
{\Gamma(a_{\{m\}}-b'_{\{m\}})\over \Gamma(-2n-b'_{\{m\}}+1)}\nonumber\\
&=&{2u\over 2N-2}\sum_{n,\{m\}}^{\infty}
{e^{-2\pi ikb'_{\{m\}}}\Gamma(a_{\{m\}}-b'_{\{m\}})
\over \Gamma(1-b'_{\{m\}})}\\
& & \hspace{4cm}\times \prod_{i\ne N-1}\left({\alpha_i\over m_i!}\right)^{m_i}
 F\left({b'_{\{m\}}\over 2},{b'_{\{m\}}+1\over 2};1;
{\La^{4N-4}\over u^{4N-4}}\right),\nonumber 
\ena
where $ b'_{\{m\}}={b_{\{m\}}\over (N-1)}-{1\over (2N-2)}$.
The analytic continuation to the region
$u^{2N-2}=s_{N-1}\sim \La^{2N-2}$
and the quadratic transformation show that the result is
\bea
a_k&=&
{2u\over 2N-2}\sum_{n,\{m\}}^{\infty}{e^{-2\pi ib'_{\{m\}}}
\Gamma(a_{\{m\}}-b'_{\{m\}})\over 
\Gamma(1-b'_{\{m\}})}
\prod_{i\ne N-1}\left({\alpha_i\over m_i!}\right)^{m_i}
\nonumber \\
&\times
 &\left[{\Gamma({1\over 2}-b'_{\{m\}})\over \Gamma(1-b'_{\{m\}})
}\left({\La^{2N-2}\over
u^{2N-2}}\right)^{-b'_{\{m\}}}\left({\La^{2N-2}+u^{2N-2}\over \La^{2N-2}}
\right)^{{1\over 2}-b'_{\{m\}}}
F\left({1\over 2},{1\over 2};b'_{\{m\}}+{1\over 2};z\right)
\right.\nonumber \\
& &\hspace{1cm}+
{\Gamma(b'_{\{m\}}-{1\over 2})\over \Gamma(b'_{\{m\}})
}(1-{\La^{4N-4}\over u^{4N-4}})^{
{1\over 2}-b'_{\{m\}}}
\left({\La^{2N-2}\over u^{2N-2}}\right)^{b'_{\{m\}}-1}
 \\
& & \hspace{4cm}\times \left.\left({\La^{2N-2}+u^{2N-2}\over \La^{2N-2}}
\right)^{b'_{\{m\}}-{1\over 2}}
F\left({1\over 2},{1\over 2};{3\over 2}-b'_{\{m\}};z\right)\right],\nonumber
\ena
where $z={1\over 2}(1-{u^{2N-2}\over \La^{2N-2}})$.

Next we consider $a_D^k\ (1\le k\le N-1)$. In this case we integrate 
meromorphic differential $\la$
 from $z=-e^{2\pi i k\over 2N-2}$ to $z=
e^{2\pi i k\over 2N-2}$ and 
evaluate douple pole of the 
integrant without 
muliplied by $\sin 2s \pi$, and subtract ${1\over 2}a_k$ \cite{MS2}. 
We have $a_D^k$ in the form: 
\bea
a_D^k&=&{u\over 2\pi^2 i}\sum_{n,\{m\}}
{e^{-2\pi ik(b'_{\{m\}})}\Gamma(a_{\{m\}}-b'_{\{m\}})\sin(b'_{\{m\}}\pi)
2^{b'_{\{m\}}}\over 
(2N-2)\Gamma({1\over 2})\Gamma(n+1)^2}
\prod_{i\ne N-1}^{N}\left({\alpha_i\over m_i!}\right)^{m_i}\nonumber \\
& &\hspace{2cm}
\times \Gamma(n+{b'_{\{m\}}\over 2})\Gamma(n+{b'_{\{m\}}\over 2}+{1\over 2})
\left({\La^{4N-4}\over u^{4N-4}}\right)^n\\
& &\times\left[\psi(n+{b'_{\{m\}}\over 2})+\psi(
n+{b'_{\{m\}}\over 2}+{1\over 2})-2\psi(n+1)+\ln\left({\La^{4N-4}\over 
u^{4N-4}}\right)+2\pi\cot(b'_{\{m\}}\pi)\right].\nonumber 
\ena
We make use of the 
analytic continuation of $a_D^k$ around the conformal point to get
\bea
a_D^k&=&
{2u\over (2N-2) i}\sum_{n,\{m\}}^{\infty}{e^{-2\pi ikb'_{\{m\}}}
\Gamma(a_{\{m\}}-b'_{\{m\}})\over \Gamma(1-b_{\{m\}})}
\prod_{i\ne N-1}\left({\alpha_i\over m_i!}\right)^{m_i}
\nonumber \\
& &\times 
\cot(b'_{\{m\}}\pi)
{\Gamma(b'_{\{m\}}-{1\over 2})\over \Gamma(b'_{\{m\}})}
(1-{\La^{4N-4}\over u^{4N-4}})^{
{1\over 2}-b'_{\{m\}}}
\left({\La^{2N-2}\over u^{2N-2}}\right)^{b'_{\{m\}}-1}
\nonumber \\
& & \hspace{4cm}\times \left({\La^{2N-2}+u^{2N-2}\over \La^{2N-2}}
\right)^{b'_{\{m\}}-{1\over 2}}
F\left({1\over 2},{1\over 2};{3\over 2}-b'_{\{m\}};z\right).\nonumber
\ena
As in the pure $SU(N)$ theory, we can claim that $a_k\sim a_D^k$ at the 
critical point. 
The behavior of $a$ and $a_D$ near $s_{N-1}=\La^{2N-2},\ s_i=0\ (i\ne N-1)$
 is 
\bea
a_k\sim a_D^k\sim (s_{N-1}-\La^{2N-2})^{{1\over 2}-{1\over 2N-2}}+\ const.
\ena
Therefore, we see that the confomal dimension of $s_{N-1}$ is 
${2N-2\over N}$ \cite{EHIY}.

As was the case of $SU(2n)$, $a_i$ and $a_D^i$ contain 
the logarithmic terms 
coming from the factor $\Gamma({1\over 2}-b'_{\{m\}})$ when $N$ of $SO(2N)$ 
is even, which vanish at the conformal point. 

Next we consider 
$a_N$ and $a_D^N$. 
Until now the calculation is same as $SU(N)$ case. 
However in order to calculate $a_N$ and $a_D^N$, we have to pick 
up the pole $x\sim 0$. To this end we rescale the variable of 
the curve as 
\bea
x^2=-{s_N\over s_{N-1}}z^2,\ \ 
\beta_i={s_i\over s_N}\left(-{s_N\over s_{N-1}}\right)
^{N-i},\ena
where $s_0=-1$, and $\la$ becomes as
\bea
\la&=&\left(-{s_N\over s_{N-1}}\right)^{1\over 2}
\int_{-i\infty}^{i\infty}{ds\over 2\pi i}
\sum_{m}{\Gamma(s+{1\over 2})\Gamma(-s)\Gamma(2s+c_{\{m\}})\over 
\Gamma({1\over 2})2s\Gamma(2s)}\left(-{\La^{4N-4}\over 
s_{N-1}^2}\right)^s\nonumber \\
& &\hspace{3cm}\times \prod_{i=0}^{N-2}\left({\beta_i^{m_i}\over m_i !}\right)
z^{4s+2Nc_{\{m\}}-2d_{\{m\}}}(z^2-1)^
{-2s-c_{\{m\}}},
\ena
where  $\{m\}=\{m_0,m_1,\cdots,m_{N-2}\}$ and $
c_{\{m\}}=\sum_{i=0}^{N-2}m_i,\ \ 
d_{\{m\}}=\sum_{i=0}^{N-2}(N-i)m_i $. 
By evaluating the line 
integral from $z=0$ to $z=1$ 
 and by multiplying $\sin 2s\pi/ \pi$ to pick up the pole at $z=1$, 
we get  
$a_N$ in the region ${s^2_{N-1}\over \La^{4N-4}}>> 1$ in the form:
\bea
a_N&=&\left(-{s_N\over s_{N-1}}\right)^{1\over 2}
\sum_{n,\{m\}}{\Gamma(n+{1\over 2})\Gamma(2n+d_{\{m\}}+
{1\over 2})\over \Gamma({1\over 2})\Gamma(2n+1)\Gamma((N-1)c_{\{m\}}-
d_{\{m\}}+{3\over 2})}\left({\La^{4N-4}\over
s_{N-1}^2}\right)^n
\prod_{i=0}^{N-2}\left({\beta_i^{m_i}\over m_i !}\right)\nonumber \\
&=&2\left(-{s_N\over s_{N-1}}\right)^{1\over 2}\sum_{\{m\}}
{\Gamma(d_{\{m\}}+
{1\over 2})\over \Gamma(-c_{\{m\}}+d_{\{m\}}+{3\over 2})}
\prod_{i=0}^{N-2}\left({\beta_i^{m_i}\over m_i !}\right)\\
& &\hspace{3cm}\times F\left({d_{\{m\}}\over 2}+{1\over 4},
{d_{\{m\}}\over 2}+{3\over 4};1;{\La^{4N-4}\over s_{N-1}^2
}\right).\nonumber
\ena
Notice that this hypergeometic function gives logarithmic term by 
analytic continuation to the region ${\La^{4N-4}\over s_{N-1}^2}\sim 1$. 
To see this, we set the variable as
\bea
y={\La^{4N-4}\over s_{N-1}^2},\ \ z={\La^{2N-2}-s_{N-1}\over 
2\La^{2N-2}},
\ena
and perform the analytic continuation to the region ${s_{N-1}\over \La^{4N-4}
}\sim 1$ as
\bea
a_N&=&\left(-{s_N\over s_{N-1}}\right)^{1\over 2}\sum_{\{m\}}
{\Gamma(d_{\{m\}}+
{1\over 2})\over \Gamma({1\over 2})\Gamma(-c_{\{m\}}+d_{\{m\}}+{3\over 2})}
\prod_{i=0}^{N-2}\left({\beta_i^{m_i}\over m_i !}\right)\nonumber \\
&\times &\left\{
\left(1-y\right)^{-d_{\{m\}}-{1\over 2}}
y^{{d_{\{m\}}\over 2}-{1\over 4}}
{\Gamma(d_{\{m\}})\over \Gamma({d_{\{m\}}\over 2}+{1\over 4})^2}
\sum_{n=0}^{d_{\{m\}}-1}{({1\over 4}-{d_{\{m\}}\over 2})_n
({1\over 4}-{d_{\{m\}}\over 2})_n
\over n!(-d_{\{m\}}+1)_n}\left(1-y\right)^n
\right.\nonumber\\
& &+{y^{-{d_{\{m\}}\over 2}-{1\over 4}}
(1-z)^{-d_{\{m\}}}
\over \Gamma({3\over 4}-{d_{\{m\}}\over 2})\Gamma({1\over 4}-
{d_{\{m\}}\over 2})
\Gamma(d_{\{m\}}+1)}\sum_{n=0}^{\infty}
{({1\over 2})_n({1\over 2})_n\over n!(d_{\{m\}}+1)_n}z^n\\
& &\times \left.\left[
\psi(n+1)+\psi(n+d_{\{m\}}+1)-2\psi(n+{1\over 2})-\pi
-\log(-z)\right]\right\}.\nonumber
\ena

Next we calculate $a_D^N$. In the region $s_{N-1}\sim \infty$, $a_D^N$ 
is given by 
integrating meromorphic differential $\la$ 
from $z=-1$ to $z=1$ without multiplying $\sin 2s\pi$ 
and subtracting ${1\over 2}a_N$, and by evaluating double poles as
\bea
a_D^N&=&{is_N^{1\over 2}\over s_{N-1}^{1\over 2}2\pi i}
\sum_{n,\{m\}}{\Gamma(d_{\{m\}}+
{1\over 2})\over \Gamma({1\over 2})\Gamma(-c_{\{m\}}+d_{\{m\}}+{3\over 2})}
\prod_{i=0}^{N-2}\left({\beta_i^{m_i}\over m_i !}\right)
{({d_{\{m\}}\over 2}+{1\over 4})_n({d_{\{m\}}\over 2}+{3\over 4})_n
\over (n!)^2}y^n
\nonumber \\
& &\hspace{1cm}\times \left[\psi(n+{d_{\{m\}}\over 2}+{1\over 4})+
\psi(n+{d_{\{m\}}\over 2}+{3\over 4})-2\psi(n+1)+
\ln y \right],
\ena
where $y={\La^{4N-4}\over s_{N-1}^2}$. Although this logarithmic term disappears by the analytic continuation 
to the region $s_{N-1}\sim \La^{2N-2}$, another logarithmic term appears  
\bea
a_D^N&=&{s_N^{1\over 2}\over 2s_{N-1}^{1\over 2}}
\sum_{n,\{m\}}{\Gamma(d_{\{m\}}+
{1\over 2})\over \Gamma({1\over 2})\Gamma(-c_{\{m\}}+d_{\{m\}}+{3\over 2})}
\prod_{i=0}^{N-2}\left({\beta_i^{m_i}\over m_i !}\right)\nonumber 
\\
&\times &\left\{
\left(1-y\right)^{-d_{\{m\}}-{1\over 2}}
y^{{d_{\{m\}}\over 2}-{1\over 4}}
{i\pi \Gamma(d_{\{m\}})\over \Gamma({d_{\{m\}}\over 2}+{1\over 4})^2}
\sum_{n=0}^{d_{\{m\}}-1}{({1\over 4}-{d_{\{m\}}\over 2})_n
({1\over 4}-{d_{\{m\}}\over 2})_n
\over n!(-d_{\{m\}}+1)_n}\left(1-y\right)^n
\right.\nonumber\\
& &+ {y^{-{d_{\{m\}}\over 2}-{1\over 4}}(1-z)^{-d_{\{m\}}}\over 
\Gamma({3\over 4}-{d_{\{m\}}\over 2})\Gamma({1\over 4}-{d_{\{m\}}\over 2})
\Gamma(d_{\{m\}}+1)}\sum_{n=0}^{\infty}{({1\over 2})_n({1\over 2})_n
z^n\over n!(d_{\{m\}}+1)_n} \\
& &\times\left. \left[\psi(n+1)+\psi(n+d_{\{m\}}+1)-2\psi(n+{1\over2 })-
\log(-z)-\pi\right]\right\}.\nonumber
\ena
Thus in $SO(2N)$ theory $a_N$ and $a_D^N$ have the 
logarighmic terms around this point though the curve 
become degenerate multiple.
Let us consider what is happening. 
Near $x\sim 0$, $\alpha_N$ cycle and $\beta_N$ cycle form 
a small torus, and the curve looks like the  curve of pure $SU(2)$ theory. 
In this case due to our choice of approaching to the point 
 $s_{N-1}= \La^{2N-2},\ 
s_N= 0$, this point 
 corresponds to the dyon point for $a_N$ and $a_D^N$ and these 
 have certainly the logarithmic terms. 
 These logarithmic terms are simply caused by the fact that we consider 
a branch where two of the singurality approach to zero before the 
theory is going to be at the critical point. 
 This point has been understood in the framework of the 
  $SU(3)$ theory near 
$u=0,\ v=\La^2$ \cite{AD}. From the expression (4.16) and (4.18), 
we see that $a_N\sim a_D^N$ on the conformal point. Therefore the existence 
of logarithmic terms in the expression (4.16) and (4.18) is not harmful.

\sect{Discussion}

We have derived the expression for the periods and Higgs fields and its 
dual around the conformal point of $SU(2)$ Yang-Mills 
theory with matter fields, pure $SU(N)$ and pure $SO(2N)$ Yang-Mills theory. 
In the $SU(2)$ theory with matter fields and the pure $SU(N)$ theory, 
we have directly recognized the structure of the theories near 
the conformal points. We find a simple correspondence between 
the fixed point of
4-D $N=2$ $SU(2)$ Yang-Mills theory with
matter fields and Landau-Ginzburg description of 2-D $N=2$
SCFT with $c=3$. 
For $SU(N)$ and $SO(2N)$ theories we could show a verification of the 
analytic continuation due to the well known formula of the 
hypergeometric functions. 

It seems interesting that we could obtain the 
explicit expression of fields around the conformal point even 
for the theories with higher rank gauge gourps.  But 
the examples we treated in this paper is elementary 
compared to more complicated 
varieties of critical points as was shown in \cite{EHIY}. 
At present, we do not know whether we can find more interesting examples
 which one can calculate the explicit form of fields. 
An important question is the verification of the validity of the analytic 
continuation for these cases, which require further investigation.

\vskip 2cm
\begin{center}
{\bf Acknowledgment}
\end{center}

We would like to thank members of the particle physics group in Hokkaido 
University for encouragement. 
One of authers (T.M.) is partially 
supported by Nukazawa Science Foundation.

\end{document}